\documentclass[preprint,nofootinbib]{revtex4}

\usepackage{graphicx}

\usepackage{amsmath}

\begin{document}

\title{On the horizon area of effective loop quantum black holes}

\author{F. C. Sobrinho$^{1,2}$, H. A. Borges$^2$, I. P. R. Baranov$^3$ and S. Carneiro\footnote{Corresponding author}\footnote{saulo.carneiro.ufba@gmail.com}$^{2,4}$}

\affiliation{$^1$Instituto de F\'isica, Universidade de S\~ao Paulo, 05508-090, S\~ao Paulo, SP, Brazil\\$^2$Instituto de F\'{\i}sica, Universidade Federal da Bahia, 40210-340, Salvador, BA, Brazil\\$^3$Instituto Federal de Educa\c c\~ao, Ci\^encia e Tecnologia da Bahia, 40301-015, Salvador, BA, Brazil\\$^4$Observat\'orio Nacional, 20921-400, Rio de Janeiro, RJ, Brazil}

\date{\today}

\begin{abstract}

Effective models of quantum black holes inspired by Loop Quantum Gravity (LQG) have had success in resolving the classical singularity with polymerisation procedures and by imposing the LQG area gap as a minimum area. The singularity is replaced by a hypersurface of transition from black to white holes, and a recent example is the Ashtekar, Olmedo and Singh (AOS) model for a Schwarzschild black hole. More recently, a one-parameter model, with equal masses for the black and white solutions, was suggested by Alonso-Bardaji, Brizuela and Vera (ABBV). An interesting feature of their quantisation is that the angular part of the metric retains its classical form and the horizon area is therefore the same as in the classical theory. In the present contribution we solve the dynamical equations derived from the ABBV effective Hamiltonian and, by applying the AOS minimal area condition, we obtain the scaling of the polymerisation parameter with the black hole mass. We then show that this effective model can also describe Planck scale black holes, and that the curvature and quantum corrections at the horizon are small even at this scale. By generating the exterior metric through a phase rotation in the dynamical variables, we also show that, for an asymptotic observer, the Kretschmann scalar is the same as in the classical Schwarzschild solution, but with a central mass screened by the quantum fluctuations.

\end{abstract}

\maketitle

\section{Introduction}

The challenge for a complete formulation of quantum gravity has, among its main goals, the resolution of the black hole and cosmological singularities as the result of space-time quantisation at the Planck scale. A promising path in this way is given by Loop Quantum Gravity (LQG), a non-perturbative quantisation of General Relativity with Ashtekar-Barbero variables, performed in the space of holonomies \cite{lewandowski,livros1,livros2,livros3}. Nevertheless, the approach to the singularity problem, both in the cosmological and black hole scenarios, has only been achieved with the help of effective models, inspired by full LQG \cite{bojwald}.

In these models the black hole physical singularity is replaced by a transition hypersurface where a black hole to white hole tunneling takes place \cite{modesto,corichi}. This is usually achieved by the following steps. First, the identification of an isometry between the black hole interior and a homogeneous background, e.g. the Kantowski-Sachs metric in the Schwarzschild case. Second, the polymerisation of the classical metric by introducing a set of parameters that control its quantum fluctuations. Different polymerisation schemes have been proposed, leading to quantum black holes with characteristic features. The dynamics is then driven by the LQG constraints, together with a minimal area postulation that constrains the evolution of the dynamical variables, giving origin to the transition surface.

A model particularly studied in recent years is that proposed by Ashtekar, Olmedo and Singh (AOS) \cite{PRL,AOS}, where a particular polymerisation scheme is adopted and the LQG minimal area is imposed on plaquettes defined by holonomies on the transition surface. In spite of giving a proper dependence of the quantum corrections on the black hole mass, which decrease as the mass increases, as well as the desired black hole to white hole transition and a smooth match between the internal and external metrics, the external solution does not present the expected asymptotic limit, as pointed out by some authors\footnote{See, however, Ref.~\cite{olmedo}.} \cite{mariam}. Other proposals try to treat this and other problems with diverse polymerisation schemes \cite{alemaes1,alemaes2,guillermo1,guillermo2}.

An interesting approach was recently proposed by Alonso-Bardaji, Brizuela and Vera (ABBV) \cite{espanhois}, where the quantisation is performed under the condition that infinitesimal coordinate transformations and gauge transformations coincide and define the same canonical algebra. The authors find in this way an internal solution with a black hole to white hole transition driven by a unique quantum parameter, and an external solution with a proper asymptotic behaviour. A remarkable feature of their quantisation scheme is that the angular part of the metric maintains its classical form, and for this reason the horizon area is the classical one, ${\cal A} = 16 \pi m^2$, where $m$ is the mass, the same for the black and white holes\footnote{In other effective models the black and white holes have in general different masses, see e.g. \cite{modesto,corichi,alemaes1,alemaes2}.}.

The latter feature is noteworthy for two reasons. First, it was previously shown that the horizon area correction in the AOS model is indeed negligible even for Planck scale black holes \cite{CQG}. For instance, for $m = 1$ (in Planck units) it follows that $\delta {\cal A}/{\cal A} \approx 10^{-3}$. The other reason is a curious coincidence between the classical horizon area of Planck scale extremal black holes and that predicted from the eigenvalues equation of the LQG area operator \cite{CQG,FoP}. Indeed, the classical horizon area of an extremal rotating black hole, for which the angular momentum is $J = m^2$, is ${\cal A} = 8 \pi J$. For the smallest admissible angular momentum, given by $\hbar$, the area can be written as
\begin{equation}
{\cal A} = 8 \pi \gamma l_p^2 \sum_1^4 \sqrt{j_i(j_i+1)}, 
\end{equation}
where $l_p$ is the Planck length, provided that $\gamma = \sqrt{3}/6$ and $j_i = 1/2$. The above equation can be identified with the eigenvalues equation of the LQG area operator \cite{rovelli} if we identify $\gamma$ with the Barbero-Immirzi parameter. It represents a horizon pierced by four lines in the fundamental representation of SU(2), and the secondary quantum numbers, for which $|m_i| = 1/2$, can be chosen in order to fulfill the projection constraint $\sum m_i = 0$. In this way we have an isolated horizon, as should be for an extremal horizon for which there is no Hawking radiation.

The value found for $\gamma$ is $5\%$ above the approximate value found from the Bekenstein-Hawking entropy of large mass horizons when we adopt the Gosh-Mitra counting of micro-states \cite{meissner,ghosh}. Furthermore, with $\gamma = \sqrt{3}/{6}$ we obtain precisely the correct leading order slope of the entropy $\times$ area relation for Planck scale black holes \cite{CQG}. This value for $\gamma$ also allows an approximate identification between the LQG minimum area and the frequency gap in the high-tone quasi-normal modes spectrum of extremal rotating black holes \cite{GRG}.

Another curious case is that of an extremal charged black hole, for which the classical horizon area is ${\cal A} = 4 \pi m^2$ and the relative correction obtained from the AOS model is $\approx 6\%$ for $m = 1$ \cite{CQG}. For a Planck mass, the horizon area corresponds to the LQG area eigenvalue of a horizon pierced by two lines with $j_i = 1/2$, which, again, can be made isolated. As $Q = m$ for extremal horizons, we see that $Q = 1$ in this case. This charge could, in principle, be identified with the elementary charge at the Planck scale, if we assume a large charge screening leading to the observed elementary charge $e \approx 0.1$ at low energies. It is also worth of note that extremal, Planck scale primordial black holes, rotating or charged, have been shown as viable candidates for composing the cosmological dark matter \cite{PLB,Nelson}.

The above correspondence between classical areas and LQG eigenvalues may, therefore, suggest that the quantum corrections to the horizon area are not only negligible, as in the AOS model, but actually null, as in the ABBV formulation. The main goal of the present paper is to explore some aspects of the latter, in particular the solution of the dynamical equations and the derivation of an explicit dependence of the polymerisation parameter on the black/white hole mass, following the AOS prescription of minimum areas on the transition surface. The Kretschmann scalar will also be computed, showing that it peaks at the transition surface and is very low at the horizon, even in the case of a Planck scale black hole. At the exterior region, with a metric derived through a phase rotation in the dynamical variables, the curvature measured by an asymptotic observer is the same as in the classical Schwarzschild solution with a screened central mass.

\section{Classical and effective Hamiltonians}

The classical Hamiltonian for the homogeneous metric is given by \cite{espanhois}
\begin{equation} \label{1}
\tilde{H}_{\text{cl}} = \dfrac{1}{G}\left[-\dfrac{\tilde{E}^{\varphi}}{2\sqrt{\tilde{E}^x}}(1+\tilde{K}_{\varphi}^2) - 2\sqrt{\tilde{E}^x}\tilde{K}_x\tilde{K}_{\varphi}\right],
\end{equation}
where $\tilde{E}^x$ and $\tilde{E}^{\varphi}$ are the components of the reduced triad and $\tilde{K}_x$ and $\tilde{K}_{\varphi}$ their conjugate momenta.
The AOS variables, on the other hand, obey the algebra $\{b, p_b\} = G\gamma$ and $\{c, p_c\} = 2G\gamma$. Therefore, if we use the substitutions $\tilde{E}^{\varphi} \rightarrow p_b$, $\tilde{E}^x \rightarrow p_c$, $\tilde{K}_{\varphi} \rightarrow b/\gamma$ and $\tilde{K}_x \rightarrow c/2\gamma$, we obtain
\begin{align}
\tilde{H}_{\text{cl}}
&= \dfrac{1}{G}\left[-\dfrac{p_b}{2\sqrt{p_c}}\left(1+\dfrac{b^2}{\gamma^2}\right) - \dfrac{\sqrt{p_c}}{\gamma^2}bc\right].\label{2}
\end{align}
Multiplying this result by the AOS lapse
\begin{equation}
N_{\text{cl}} = \dfrac{\gamma}{b}\sqrt{p_c},
\end{equation}
we have
\begin{align}
\tilde{H}_{\text{cl}}[N_{\text{cl}}]
&=-\dfrac{1}{2G\gamma}\left[2cp_c + \left(b + \dfrac{\gamma^2}{b}\right)p_b\right]\label{3}\\
&= \tilde{H}_{\text{cl}}^{\text{AOS}}[N_{\text{cl}}]\nonumber.
\end{align}
This expresses the classical Hamiltonian in terms of the variables used in the AOS paper. Now, we perform a polymerisation by substituting\footnote{For a discussion on the covariance of this and other polymerisation schemes, see \cite{bojowald}.} \cite{florencia}
\begin{equation}\label{4}
b \rightarrow \dfrac{\sin\left(\delta_b b\right)}{\delta_b}, \quad \quad \quad p_b \rightarrow \dfrac{p_b}{\cos(\delta_b b)},
\end{equation}
and by including the regularisation factor
\begin{equation}\label{5}
\dfrac{\cos(\delta_b b)}{\sqrt{1+\gamma^2\delta_b^2}}.
\end{equation}
This results in the effective Hamiltonian
\begin{align}
H_{\rm eff}[N_{\rm eff}]
&= -\dfrac{1}{2G\gamma\sqrt{1+\gamma^2\delta_b^2}}\left[2cp_c\cos(\delta_b b) + \left(\dfrac{\sin(\delta_b b)}{\delta_b} + \dfrac{\delta_b\gamma^2}{\sin(\delta_b b)}\right)p_b\right].\label{6}
\end{align}

On the other hand, if we first multiply (\ref{1}) by the AOS classical lapse in variables of extrinsic curvature, 
\begin{equation}\label{7}
N_{\text{cl}} = \dfrac{\sqrt{\tilde{E}^x}}{\tilde{K}_{\varphi}},
\end{equation}
we obtain
\begin{equation}\label{8}
\tilde{H}_{\text{cl}}[N_{\text{cl}}] = \dfrac{1}{G}\left[-\dfrac{\tilde{E}^{\varphi}}{2}\left(\dfrac{1}{\tilde{K}_{\varphi}}+\tilde{K}_{\varphi}\right) - 2\tilde{E}^x\tilde{K}_x\right].
\end{equation}
Now, by performing the polymerisation in the ABBV form \cite{espanhois}
\begin{equation}\label{9}
    \tilde{K}_{\varphi} \rightarrow \dfrac{\sin(\lambda K_{\varphi})}{\lambda}, \quad \quad \quad \tilde{E}^{\varphi} \rightarrow \dfrac{E^{\varphi}}{\cos(\lambda K_{\varphi})},
\end{equation}
we find
\begin{equation}\label{10}
H_{\rm eff}[N_{\rm eff}] = \dfrac{1}{G}\left[-\dfrac{E^{\varphi}}{2\cos(\lambda K_{\varphi})}\left(\dfrac{\lambda}{\sin(\lambda K_{\varphi})}+\dfrac{\sin(\lambda K_{\varphi})}{\lambda}\right) - 2E^xK_x\right].
\end{equation}
We then multiply it by the regularisation factor
\begin{equation}\label{11}
   \dfrac{\cos(\lambda K_{\varphi})}{\sqrt{1+\lambda^2}}, 
\end{equation}
to obtain
\begin{eqnarray}\label{12}
H_{\rm eff}[N_{\rm eff}] &=& -\dfrac{1}{G\sqrt{1+\lambda^2}}\bigg[\dfrac{E^{\varphi}}{2}\left(\dfrac{\lambda}{\sin(\lambda K_{\varphi})}+\dfrac{\sin(\lambda K_{\varphi})}{\lambda}\right) \nonumber \\ &+& 2E^xK_x\cos(\lambda K_{\varphi})\bigg].
\end{eqnarray}
Finally, substituting $E^{\varphi} \rightarrow p_b$, $E^x \rightarrow p_c$, $K_{\varphi} \rightarrow b/\gamma$ and $K_x \rightarrow c/2\gamma$, we find
\begin{align}
H_{\rm eff}[N_{\rm eff}]
&= -\dfrac{1}{2G\gamma\sqrt{1+\lambda^2}}\left[2cp_c\cos\left( \dfrac{\lambda b}{\gamma}\right) + \left(\dfrac{\gamma\sin\left( \dfrac{\lambda b}{\gamma}\right)}{\lambda} + \dfrac{\lambda \gamma}{\sin\left( \dfrac{\lambda b}{\gamma}\right)}\right)p_b\right].\label{13}
\end{align}
Comparing the terms inside brackets in Eqs. (\ref{6}) and (\ref{13}), we see that the two expressions are identical if $\lambda = \gamma\delta_b$. The classical Hamiltonian is recovered for $\delta_b \rightarrow 0$.

\section{Dynamical equations and solutions}

The dynamical equations are derived as usually from the Hamiltonian formalism and are written as
\begin{align}
\dot{b} &= \{b, H_{\rm eff}\} = G\gamma\dfrac{\partial H_{\rm eff}}{\partial p_b} = -\dfrac{1}{2\sqrt{1+\gamma^2\delta_b^2}}\left(\dfrac{\sin(\delta_b b)}{\delta_b} + \dfrac{\delta_b\gamma^2}{\sin(\delta_b b)}\right)\label{14},\\
\dot{c} &= 2G\gamma\dfrac{\partial H_{\rm eff}}{\partial p_c} = -\dfrac{2c\cos(\delta_b b)}{\sqrt{1+\gamma^2\delta_b^2}}\label{15},\\
\dot{p_b} &= -G\gamma\dfrac{\partial H_{\rm eff}}{\partial b}
\nonumber
\\
&= \dfrac{1}{2\sqrt{1+\gamma^2\delta_b^2}}\left[-2cp_c\sin(\delta_b b)\delta_b + \left(1 - \dfrac{\delta_b^2\gamma^2}{\sin^2(\delta_b b)}\right)p_b\cos(\delta_b b)\right]\label{16},\\
\dot{p_c} &= -2G\gamma\dfrac{\partial H_{\rm eff}}{\partial c} = \dfrac{2p_c\cos(\delta_b b)}{\sqrt{1+\gamma^2\delta_b^2}},\label{17}
\end{align}
where the dot means derivative with respect to the AOS time variable $T$.

The solution for $b(T)$ is given by\footnote{The integration constant was chosen so that the horizon corresponds to $T = 0$, when $\cos(\delta_b b) = 1$ and, in metric (\ref{metric_interior}), the lapse diverges and $p_b$ vanishes.}
\begin{equation}\label{18}
\cos(\delta_b b) = b_0\left[\dfrac{1 + b_0\tanh\left(\frac{T}{2}\right)}{b_0 + \tanh\left(\frac{T}{2}\right)}\right],
\end{equation}
with
\begin{equation} \label{b0}
b_0 = \sqrt{1 + \gamma^2\delta_b^2}.
\end{equation}
Using (\ref{18}) into (\ref{15}) and (\ref{17}) we have, after integration,
\begin{align}
c(T) &= c^{(0)} \left[b_0\cosh\left(\dfrac{T}{2}\right) + \sinh\left(\dfrac{T}{2}\right)\right]^{-4}\label{19}, \\
p_c(T) &= p_{c}^{(0)} \left[b_0\cosh\left(\dfrac{T}{2}\right) + \sinh\left(\dfrac{T}{2}\right)\right]^{4}.\label{20}
\end{align}
The solution for (\ref{16}) can be obtained from the Hamiltonian constraint $H_{\rm eff} = 0$,
\begin{equation}\label{21}
p_b(T) = -\dfrac{2 c p_c}{\dfrac{\tan(\delta_b b)}{\delta_b} + \dfrac{\gamma^2\delta_b}{\sin(\delta_b b)\cos(\delta_b b)}}.
\end{equation}

The minimum value for $p_c(T)$ is found from $\dot{p}_c = 0$,
\begin{align}
    \sinh(T_{\mathcal{T}}/2) &= -\dfrac{1}{\gamma \delta_b}\label{22},\\
        p_{c}^{\text{min}} = p_{c}(T_\mathcal{T}) &= p_c^{(0)}\gamma^4\delta_b^4,\label{23}
\end{align}
with $T_{\mathcal{T}}$ corresponding to the transition surface. At the horizon we have
\begin{equation}\label{24}
    p_c^{\text{hor}} = p_c^{(0)} b_0^4.
\end{equation}
It is worth of note that, from (\ref{23}) and (\ref{24}), we always have $p_c^{\text{min}} < p_c^{\text{hor}}$, that is, the transition surface is inside the horizon whatever the value of $\delta_b$, which is not generally the case in the AOS model \cite{CQG}.

The value of $p_c^{(0)}$ can be expressed in terms of the invariant mass \cite{espanhois}
\begin{equation}\label{invariant mass}
    m = \frac{\sqrt{p_c}}{2} \left( 1 + \frac{\sin^2(\delta_b b)}{\gamma^2 \delta_b^2} \right).
\end{equation}
At the horizon we have, from (\ref{18}), $\sin^2(\delta_b b) = 0$, i.e.
\begin{equation} \label{pc0}
    p_c^{(0)} = \frac{4m^2}{b_0^4}.
\end{equation}
Therefore, the horizon area is given by
\begin{align}
    {\cal A} &= 4\pi p_c^{\text{hor}} = 16\pi m^2,\label{26}
\end{align}
that is, the same classical area, as expected since the angular term of the metric is unperturbed in the adopted polymerisation.

\section{Planck scale black holes}

In effective models the quantum corrections are controlled by the polimerisation parameters ($\delta_b$ in the present case). From a physical viewpoint, we expect that such corrections decrease for larger black holes, vanishing in the classical limit. For example, in the present model the radius of the transition surface is determined by (\ref{23}) and vanishes for $\delta_b \rightarrow 0$, when we recover the classical singularity. The procedure adopted in the AOS model for establishing the relation between the polimerisation parameters and the black hole mass has been criticised for the following reason \cite{referee2,guillermo3}. When deriving the equations of motion from the effective Hamiltonian, those parameters are considered constant on the phase space, while the invariant mass is constant along a given dynamical trajectory, but evidently depends on the phase space variables (see (\ref{invariant mass}), for example).

A possible way to formally circumvent this loophole is to enlarge the phase space, treating the polimerisation parameters as additional conjugate variables that are constant along a dynamical trajectory \cite{AOS}. In this way, the relation between $\delta_b$ and $m$ to be found in this section can be thought as a constraint that selects, from the physical trajectories of the effective Hamiltonian, those that match the full LQG area gap on the transition surface. Despite this, we are aware of the controversy surrounding this procedure\footnote{In any case, the results of the next sections do not depend on the constraint (\ref{28}), which will only be used to fix $m$ and $\delta_b$ in the figures.}.

In order to determine the dependence of the quantum parameter $\delta_b$ on the black hole mass, we follow the AOS procedure of imposing the LQG area gap as the minimum area for any plaquette defined by holonomies on the transition surface \cite{AOS}. Since in the ABBV scheme the radial coordinate is not polimerised, these minimal plaquettes can only be defined on $\theta$-$\varphi$ $2$-surfaces, and the AOS constraint is written as
\begin{equation}
    4\pi p_c^{\text{min}} (\alpha\delta_b)^2 = 4\pi \sqrt{3}\gamma\quad (l_p = 1),\label{27}
\end{equation}
where $\alpha$ is a positive parameter that defines the minimum length of the plaquette links as proportional to the polymerisation parameter $\delta_b$ \cite{CQG}. Using (\ref{b0}), (\ref{23}) and (\ref{pc0}), we obtain
\begin{equation}
    \frac{\delta_b^6}{(1 + \gamma^2 \delta_b^2)^2} = \dfrac{\sqrt{3}}{4\gamma^3(m\alpha)^2}.\label{28}
\end{equation}
As $4\pi p_c^{\text{min}} \geq 4\pi\sqrt{3}\gamma$, we also have the constraint
\begin{equation}\label{29}
    \alpha\delta_b\leq 1.
\end{equation}

\begin{figure}
\centering
\includegraphics[width=0.5\linewidth]{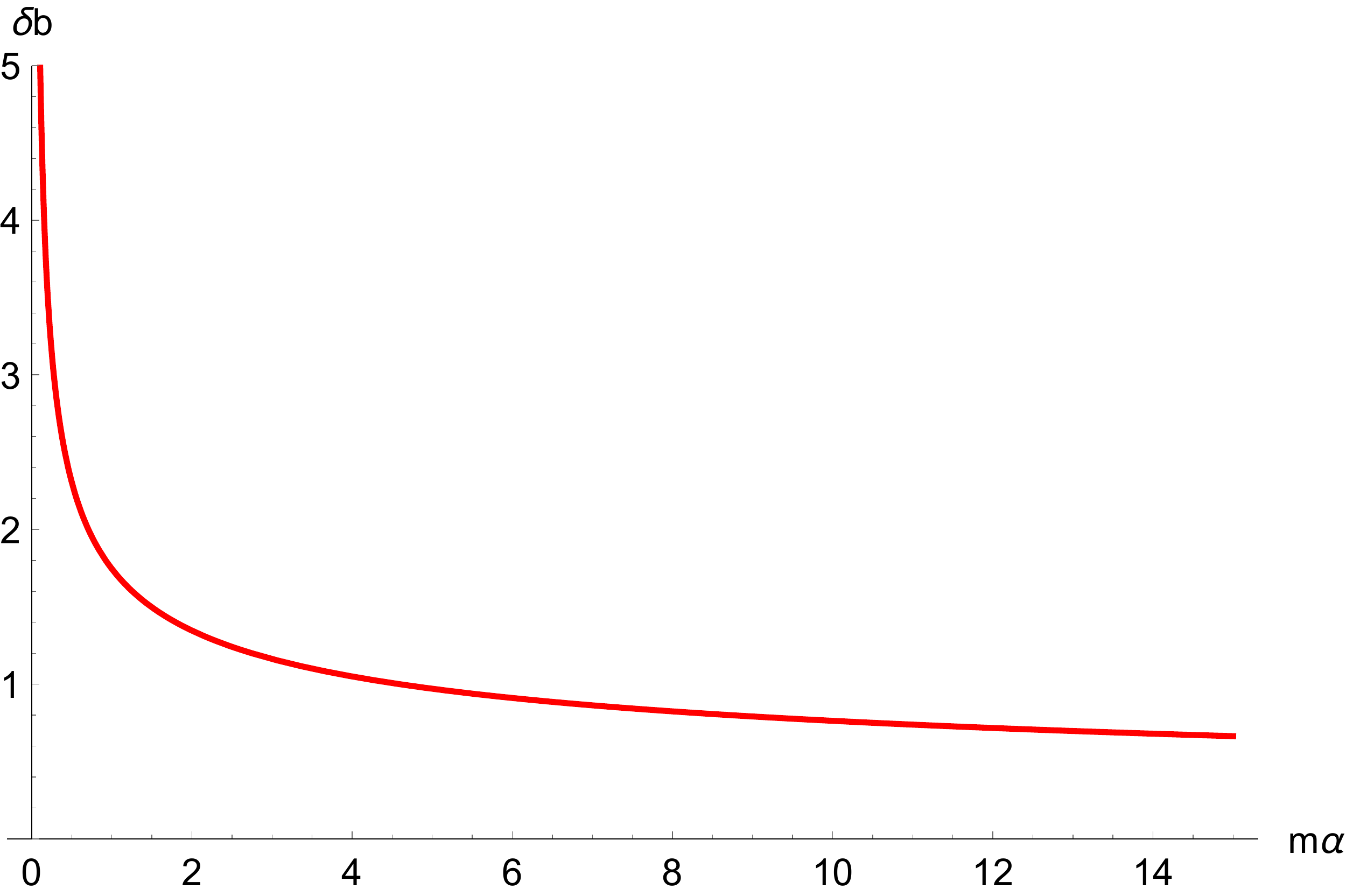}
\caption{$\delta_b$  $\times$ 
 black hole mass dependence}
\label{fig:humberto}
\end{figure}

From (\ref{28}) and Fig.~\ref{fig:humberto} we see that the quantum corrections vanish when $m \rightarrow \infty$. On the other hand, as $\delta_b$ increases when $m \rightarrow 0$, for a given value of $\alpha$ there is a minimum allowed mass, that saturates (\ref{29}). If this minimum mass has the Planck scale\footnote{In full LQG, an isolated horizon of area $16 \pi m^2$ has minimal mass $m_{\text{min}} \approx 0.5$, for $\gamma = \sqrt{3}/6$.}, $m_{\text{min}} \approx 1$, the maximum $\delta_b$, for $\gamma = \sqrt{3}/6$, is
$\delta^{\text{max}}_b \approx 2.6$, and we have $\alpha \approx 0.4$. 
On the other hand, if we fix $\alpha = 1$, as in the AOS paper, $\delta_b \leq 1$ and the minimum mass is $m_{\text{min}}\approx 4.5\, m_p$.

An absolute lower bound for the black hole mass is obtained when the transition surface approaches the horizon. From (\ref{23}) and (\ref{24}) this corresponds to the limit $\gamma \delta_b \gg 1$. Hence, from (\ref{28}) we have $\alpha^2 \delta_b^2 = \sqrt{3}\gamma/(2m)^2$, which saturates (\ref{29}) for $m^2 = \sqrt{3}\gamma/4$. Incidentally, this is the minimum mass allowed in full LQG for a horizon of area $16 \pi m^2$, corresponding to a horizon pierced by a single spin network line with $j = 1/2$. This bound is actually never reached, since we always have $p_c^{\text{min}} < p_c^{\text{hor}}$.

\section{Curvature invariants}

The polymerisation removes the singularity present in the classical metric, and this can be evidenced by calculating some curvature invariants as, for instance, the Kretschmann scalar. At the same time, we can use it to evaluate the effect of the quantum corrections at the horizon.
The Kretschmann scalar is a combination of several terms and can be calculated using the homogeneous metric,
which, after polimerisation, can be written in AOS variables as
\begin{equation} \label{metric_interior}
ds^2 = -N^2 dT^2 + \frac{p_b^2}{p_c \cos^2(\delta_b b)} dx^2 + p_c d\Omega^2,
\end{equation}
with the lapse given by
\begin{equation} \label{lapse_interior}
N^2 = \frac{\gamma^2 \delta_b^2 p_c}{\sin^2(\delta_b b)}.
\end{equation}
Using the solutions of Sec.~III, it is possible to rewrite
metric (\ref{metric_interior}) in the ABBV form \cite{espanhois}
\begin{equation} \label{ABBV interior}
ds^2 =  - \left( \frac{2m}{\tilde{r}} -1 \right)^{-1} \left( 1 - \frac{r_0}{\tilde{r}} \right)^{-1} d\tilde{r}^2 + \left( \frac{2m}{\tilde{r}} - 1 \right) d\tau^2 + \tilde{r}^2 d\Omega^2,
\end{equation}
where we defined
\begin{equation}
\tau = \frac{c^{(0)}p_c^{(0)}}{m\gamma}\, x, \quad \quad 
\tilde{r} = \sqrt{p_c}, \label{definition1}
\end{equation}
\begin{equation}
r_0 = \sqrt{p_{c}^{\text{min}}} = \frac{2m \gamma^2 \delta_b^2}{b_0^2}. \label{definition2}
\end{equation}

\begin{figure}
\centering
\includegraphics[width=0.5\linewidth]{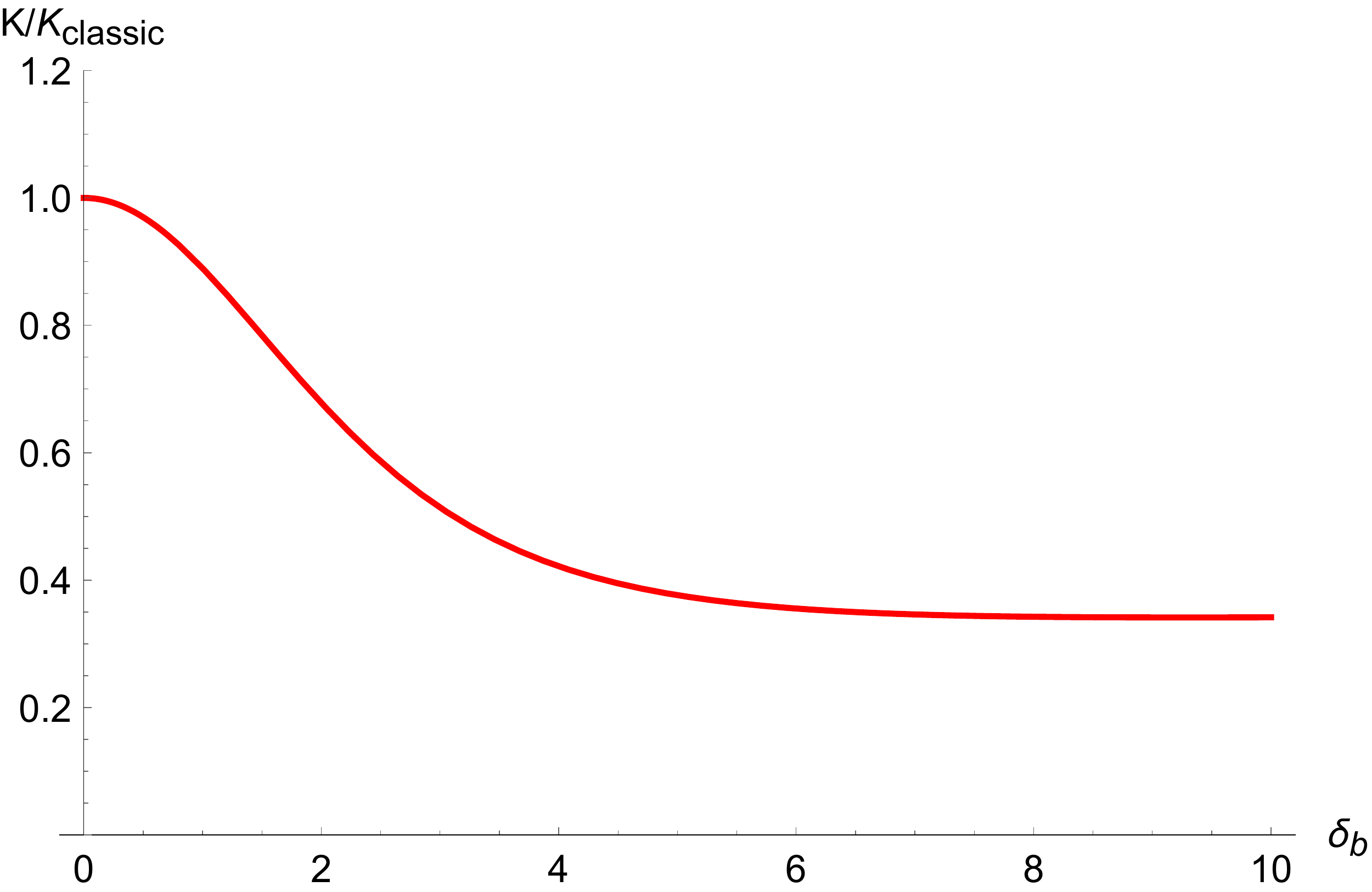}
\caption{$K/K_{\textrm{classic}}$ at the horizon as a function of $\delta_b$}
\label{fig:kporkclassic}
\end{figure}

The Kretschmann expression at the horizon $(T=0)$ is\footnote{We acknowledge the use of the \texttt{MATHEMATICA} package \texttt{xAct} in our computations: http://www.xact.es.}
\begin{equation}
    K_{\textrm{horizon}}=\frac{48 + 24 \gamma^2 \delta_b^2 + 17 \gamma^4 \delta_b^4}{64 m^4 (1 \
+ \gamma^2 \delta_b^2)^2}.
\end{equation}
For small quantum corrections, $\delta_b \rightarrow 0$, the classical result $K_{\textrm{classic}} = 3/(4 m^4)$ is recovered.
In Fig.~\ref{fig:kporkclassic} we show the behaviour of $K/K_{\textrm{classic}}$ at the horizon as a function of $\delta_b$, for $\gamma = \sqrt{3}/6$, which shows that the curvature does not deviate considerably from its classical level whatever the value of $\delta_b$.

\begin{figure}
\centering
\includegraphics[width=0.5\linewidth]{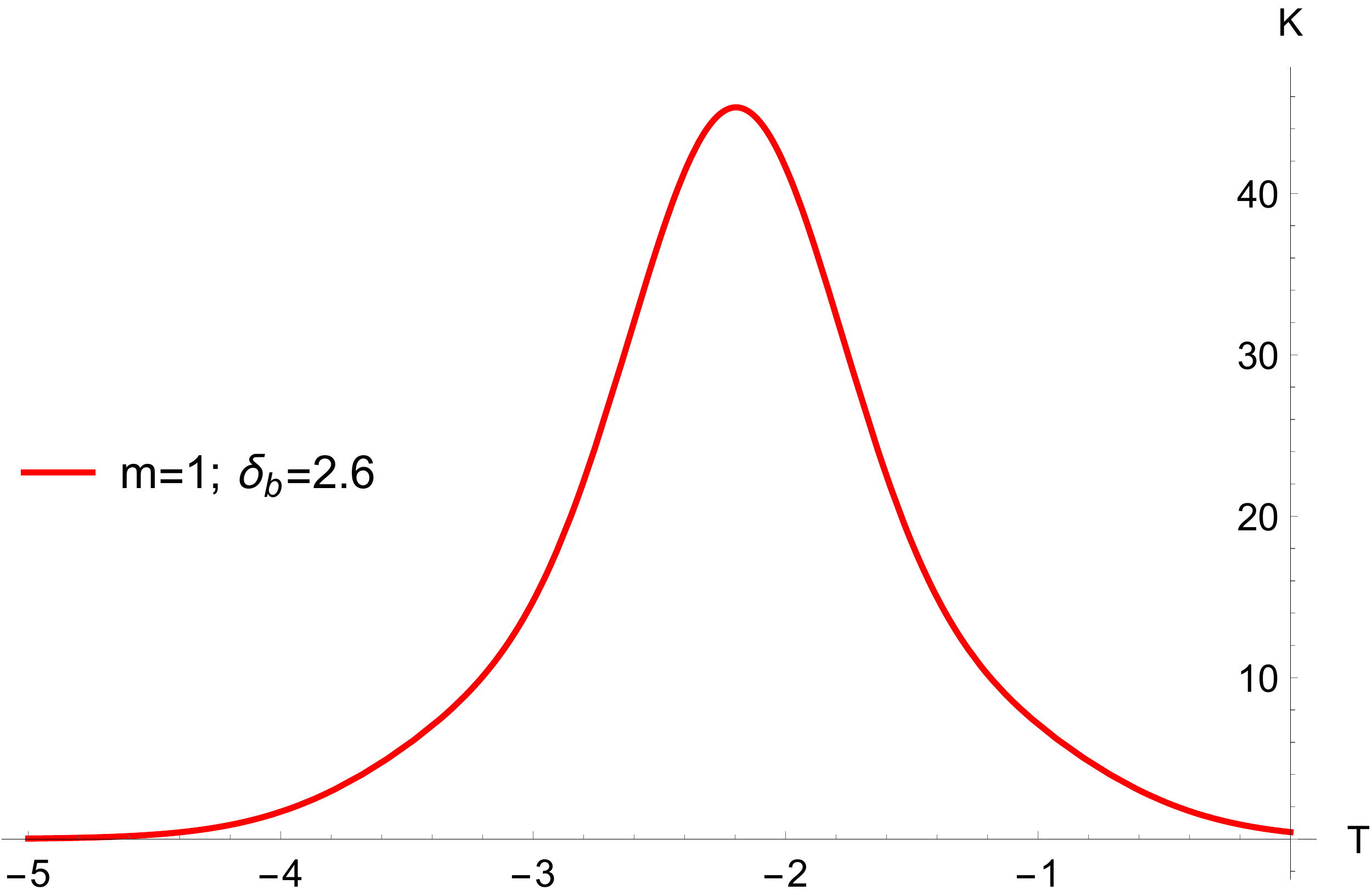}
\caption{The Kretschmann scalar as a function of $T$ for the homogeneous solution}
\label{fig:KvsT}
\end{figure}

In Fig.~\ref{fig:KvsT} we plot the Kretschmann scalar as a function of $T$, for $m=1$ and $\delta_b=2.6$. As can be seen, it has no singularity, presents a maximum at the transition surface and is very small at the black hole horizon $T=0$. Note the symmetry between the black hole and white hole phases. From (\ref{22}), the transition surface in this figure occurs at $T_{\text{min}} \approx -2.2$, while the white hole horizon corresponds to $T \approx -4.4$.
At the transition surface the  Kretschmann scalar is generally given by
\begin{equation}
    K(T_{\mathcal{T}})=\frac{(1 + \gamma^2 \delta_b^2)^4 (9 + 2 \gamma^2 \delta_b^2 + 17 \gamma^4 \delta_b^4)}{64 m^4 \gamma^{12} \delta_b^{12}}.
\end{equation}

For completeness, we obtained the Ricci scalar $(g_{\mu \nu}R^{\mu \nu})$, the square of the Ricci tensor $(R_{\mu \nu}R^{\mu \nu})$ and the Weyl scalar at the horizon and the transition surface:
\begin{equation}
   g_{\mu \nu}R^{\mu \nu}\vert_{\textrm{horizon}} = \frac{3 \gamma^2 \delta_b^2}{8 m^2 (1 + \gamma^2 \delta_b^2)},
\end{equation}
\begin{equation}
   g_{\mu \nu}R^{\mu \nu}\vert_{T_\mathcal{T}} = \frac{3 (1 + \gamma^2 \delta_b^2)^3}{8 m^2 \gamma^6 \delta_b^6},
\end{equation}
\begin{equation}
 R_{\mu \nu}R^{\mu \nu}\vert_{\textrm{horizon}} = \frac{17 \gamma^4 \delta_b^4}{128 m^4 (1 + \gamma^2 \delta_b^2)^2},   
\end{equation}
\begin{equation}
 R_{\mu \nu}R^{\mu \nu}\vert_{ T_\mathcal{T}}  = \frac{(1 + \gamma^2 \delta_b^2)^4 (9 + 14 \gamma^2 \delta_b^2 + 17 \gamma^4 \delta_b^4)}{128 m^4 \gamma^{12} \delta_b^{12}},
   \end{equation}
\begin{equation}
  C_{\alpha \beta \gamma \delta }C^{\alpha \beta \gamma \delta }\vert_{ \textrm{horizon}} = \frac{3 (4 + \gamma^2 \delta_b^2)^2}{64 m^4 (1 + \gamma^2 \delta_b^2)^2},  
\end{equation}
\begin{equation}
  C_{\alpha \beta \gamma \delta }C^{\alpha \beta \gamma \delta }\vert_{ T_\mathcal{T}}  = \frac{3 (-1 + \gamma^2 \delta_b^2)^2 (1 + \gamma^2 \delta_b^2)^4}{64 m^4 \gamma^{12} \delta_b^{12}}.
\end{equation}
At the horizon all these scalars reduce to the classical ones for $\delta_b = 0$. At the transition surface, all of them diverge for $\delta_b \rightarrow 0$, as expected.

\section{The asymptotic limit}

\begin{figure}
\centering
\includegraphics[width=0.5\linewidth]{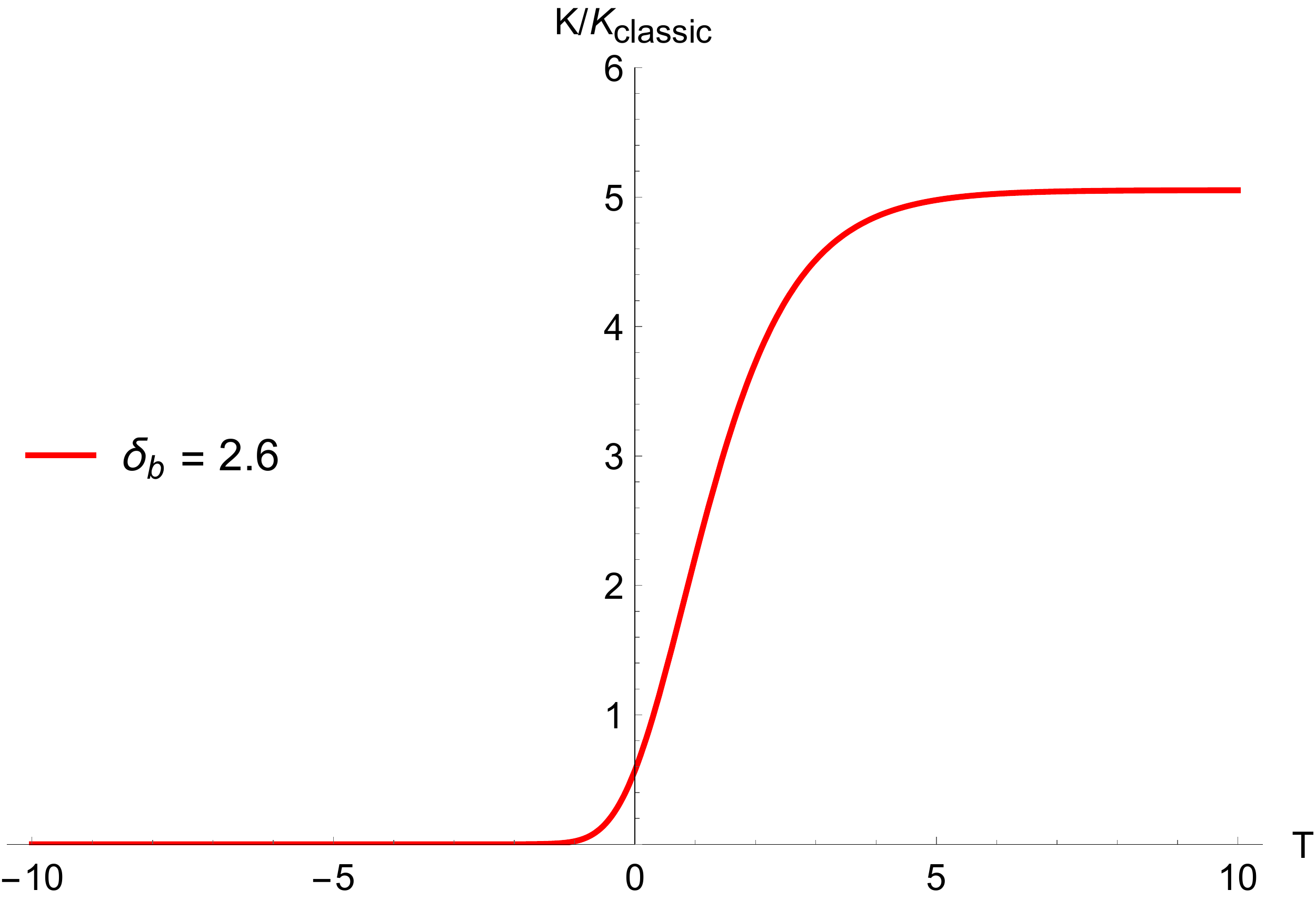}
\caption{$K/K_{\textrm{classic}}$ as a function of $T$ for $m=1$ and $\alpha = 0.4$}
\label{fig:kporkclassic2}
\end{figure}

Let us finish this discussion on the ABBV model with an analysis of the asymptotic limit for Planck scale black holes. 
While in the ABBV paper the homogeneous and static metrics are derived from the same polimerised Hamiltonian (\ref{12}) in different gauges, we will generate the exterior metric from the homogeneous one through the substitutions $b \rightarrow ib$ and $p_b \rightarrow ip_b$, which relates the interior and exterior classical metrics \cite{AOS}. From (\ref{metric_interior}) and (\ref{lapse_interior}) this leads to
\begin{equation}\label{exterior}
ds^2 = N^2 dT^2 - \frac{p_b^2}{p_c \cosh^2(\delta_b b)} dx^2 + p_c d\Omega^2,
\end{equation}
where now
\begin{equation}
N^2 = \frac{\gamma^2 \delta_b^2 p_c}{\sinh^2(\delta_b b)}.
\end{equation}

The solutions for $c$ and $p_c$ remains the same as in Sec.~III, while $b$ and $p_b$ are now determined from
\begin{equation}\label{18'}
\cosh(\delta_b b) = b_0\left[\dfrac{1 + b_0\tanh\left(\frac{T}{2}\right)}{b_0 + \tanh\left(\frac{T}{2}\right)}\right],
\end{equation}
\begin{equation}\label{21'}
p_b(T) = \dfrac{2 c p_c}{\dfrac{\tanh(\delta_b b)}{\delta_b} - \dfrac{\gamma^2\delta_b}{\sinh(\delta_b b)\cosh(\delta_b b)}}.
\end{equation}
On the other hand, the invariant mass (\ref{invariant mass}) acquires the form
\begin{equation}
    m = \frac{\sqrt{p_c}}{2} \left( 1 - \frac{\sinh^2(\delta_b b)}{\gamma^2 \delta_b^2} \right),
\end{equation}
which is identically satisfied by (\ref{20}) and (\ref{18'}). 

With these solutions, it is possible to show that 
metric (\ref{exterior}) can be written in the ABBV form \cite{espanhois}
\begin{equation}
ds^2 = - \left( 1 - \frac{2m}{\tilde{r}} \right) d\tau^2 + \left( 1 - \frac{2m}{\tilde{r}} \right)^{-1} \left( 1 - \frac{r_0}{\tilde{r}} \right)^{-1} d\tilde{r}^2 + \tilde{r}^2 d\Omega^2,
\end{equation}
where we used again the definitions (\ref{definition1}) and (\ref{definition2}). A comparison with (\ref{ABBV interior}) shows that the exterior metric can also be obtained from the interior one through analytical continuation.
This metric does not represent a Schwarzschild spacetime sourced by a central mass $m$, except for $\delta_b \ll 1$ [i.e., $m \gg 1$ in view of (\ref{28})].
Nevertheless, in Fig.~\ref{fig:kporkclassic2} we show the ratio $K/K_{\textrm{classic}}$ as a function of $T$ for both the homogeneous and static regions. For $T \rightarrow -\infty$ it goes to zero since $K_{\textrm{classic}}$ diverges at the origin.
For $T \rightarrow \infty$ it tends to a constant, equal to $1$ for $m \rightarrow \infty$ and to $5.05$ for $m = 1$. Therefore, for an asymptotic observer the curvature is the same as in the classical Schwarzschild solution, but with an effective central mass screened by the quantum fluctuations. As $K_{\textrm{classic}} = 3 e^{-6T}/4 m^4$, for $m = 1$ the screened mass is $m_{\textrm{eff}} \approx 0.67$. That the mass measured from infinity does not equal half of the Schwarzschild radius (given by $2m$ from (\ref{26})) is a general feature of polymer black holes \cite{alemaes2}.\footnote{Note, however, that the effective mass defined here does not coincide with the ADM mass, given by $M_{\text{ADM}} = m + r_0/2$ \cite{bascos2}.}

\section{Concluding remarks}

After three decades of development, Loop Quantum Gravity is nowadays a solid theory from the foundational and mathematical point of views, at least solid enough to be considered one of the best candidates for a quantum gravity theory, in spite of some loopholes that still await for a satisfactory treatment \cite{livros1,livros2,livros3}. At the same time, LQG-inspired effective models have been successful in resolving classical singularities, both at the cosmological and black hole contexts. On the other hand, finding observational signatures of space-time quantisation or some phenomenological prediction of such models also constitutes important challenge.

Recently, the possibility that Planck scale primordial and stable black holes were formed after inflation, composing today the cosmological dark matter, has been explored \cite{PLB,Nelson,referee1}. It is unlikely that this could be verified in the near future, but it certainly deserves some interest as a theoretical proposal. On the other hand, its realisation is based on the assumption that the classical area of extremal horizons remains unchanged at the quantum level. This was indeed shown to be a good approximation in the realm of the AOS model \cite{CQG}. The coincidence between classical horizon areas of Planck scale extremal black holes and eigenvalues of the LQG area operator also corroborates this assumption.

In the present paper we have analysed a recently proposed effective model where the horizon area of a spherically symmetric black hole does not suffer any quantum correction, maintaining its classical dependence on the black hole mass. Our main goal was twofold. First, an explicit derivation and solution of the dynamical equations from the ABBV effective Hamiltonian in AOS variables (see also \cite{bascos2}). Second, the use of the solutions found to evaluate the model at the Planck scale, in particular the minimal allowed mass and the quantum corrections at the horizon. The AOS model was originally proposed for the resolution of the singularity of macroscopic black holes, where the authors used an approximate solution for the dependence of the polymerisation parameters on the black hole mass. In Ref.~\cite{CQG} an exact solution for these parameters was found, which permitted the model extension to the Planck scale. 

This was also possible in the ABBV case, reinforcing the potential of effective models to mimic the main features of full LQG at so short scales \cite{Rovelli}. With the proportionality parameter $\alpha$ of the order of unity, it was possible to describe black holes of Planck mass. Even adopting $\alpha = 1$ as in the original AOS proposal, the allowed mass can be as small as four Planck masses. We have also estimated some curvature scalars at the horizon and at the black hole to white hole transition surface. They do not diverge anywhere, with the Kretschmann scalar presenting a maximum at the transition surface. Interesting enough, this scalar is comparatively negligible at the horizon even for a Planck scale mass, corroborating again the possibility of using this and other effective models at this scale. Finally, the exterior metric was derived through a phase rotation in the dynamical variables. By computing the Kretschmann scalar we have verified that, asymptotically, it presents the classical Schwarzschild form, with a central mass screened by quantum fluctuations.

The possibility of describing Planck scale black holes in the context of effective models may seem curious in view of the common belief that they are not valid approximations in the realm of high quantum corrections. Nevertheless, let us remind that, even in the case of large black holes, quantum fluctuations are large at the transition surface, whose existence is established with the help of effective models. In this sense, the potential of such models for describing microscopic black holes should not sound so surprising. Anyway, the discussion of Sec.~IV suggests an inferior mass limit for the validity of the present model, of the order of the Planck mass. The Planck scale remains therefore a frontier beyond which a full quantum gravity approach is unavoidable. 

A similar comment is in order on the quantum corrections at the horizon. From Figs.~2 and 4 we see that, for a Planck mass black hole, the Kretschmann scalar at the horizon is $\approx 40\%$ lower than in the classical solution. Although significant, this difference is not so large as one would expect for microscopic horizons. On the other hand, from Fig.~3 we see that the curvature on the horizon is negligible when compared to that at the transition surface. As the Schwarzschild radius is equal to $2m$, this suggests again that quantum fluctuations are only important at trans-Planckian scales.

\section*{Acknowledgements}

We are thankful to David Brizuela and Matheus Mello for useful discussions, and to Alberto Saa for a first reading. FCS was supported by PIBIC/CNPq (Brazil). SC is partially supported by CNPq with grant 311584/2020-9.


\begin{thebibliography}{}

\bibitem{lewandowski} A. Ashtekar and J. Lewandowski, Class. Quantum Grav. {\bf 21} (2004) R53.

\bibitem{livros1} T. Thiemann, {\it Modern Canonical Quantum General Relativity} (Cambridge University Press, 2008).

\bibitem{livros2} R. Gambini and J. Pullin, {\it A first course in Loop Quantum Gravity} (Oxford University Press, 2011).

\bibitem{livros3} C. Rovelli and F. Vidotto, {\it Covariant Loop Quantum Gravity} (Cambridge University Press, 2015).

\bibitem{bojwald} M. Bojowald, Living Rev. Relativ. {\bf 11} (2008) 4.

\bibitem{modesto} L. Modesto and I. Pr\'emont-Schwarz, Phys. Rev. {\bf D80} (2009) 064041.

\bibitem{corichi} A. Corichi and P. Singh, Class. Quantum Grav. {\bf 33} (2016) 055006.

\bibitem{PRL} A. Ashtekar, J. Olmedo and P. Singh, Phys. Rev. Lett. {\bf 121} (2018) 241301.

\bibitem{AOS} A. Ashtekar, J. Olmedo and P. Singh, Phys. Rev.
{\bf D98} (2018) 126003.

\bibitem{olmedo} A. Ashtekar and J. Olmedo, Int. J. Mod. Phys. {\bf D29} (2020) 2050076.

\bibitem{mariam} M. Bouhmadi-L\'opez {\it et al.}, Phys. Dark Univ. {\bf 30} (2020) 100701.

\bibitem{alemaes1} N. Bodendorfer, F. M. Mele and J. M\"unch, Class. Quantum Grav. {\bf 36} (2019) 195015.

\bibitem{alemaes2} N. Bodendorfer, F. M. Mele and J. M\"unch, Class. Quantum Grav. {\bf 38} (2021) 095002.

\bibitem{guillermo1} B. Elizaga Navascués, A. García-Quismondo and G. A. Mena Marugán, Phys.Rev. {\bf D106} (2022) 063516.

\bibitem{guillermo2} B. Elizaga Navascués, A. García-Quismondo and G. A. Mena Marugán, Phys.Rev. {\bf D106} (2022) 043531.

\bibitem{espanhois} A. Alonso-Bardaji, D. Brizuela and R. Vera, Phys. Lett. {\bf B829} (2022) 137075.

\bibitem{CQG} C. Pigozzo, F. S. Bacelar and S. Carneiro, Class. Quantum Grav. {\bf 38} (2021) 045001.

\bibitem{FoP} S. Carneiro, Found. Phys. {\bf 50} (2020) 1376.

\bibitem{rovelli} C. Rovelli and L. Smolin, 1995 Nucl. Phys. {\bf B442} (1995) 593;
Nucl. Phys. {\bf B456} (1995) 753 (erratum).

\bibitem{meissner} K. A. Meissner, Class. Quantum Grav. {\bf 21} (2004) 5245.

\bibitem{ghosh} A. Ghosh and P. Mitra, Phys. Lett. {\bf B616} (2005) 114.

\bibitem{GRG} S. Carneiro and C. Pigozzo, Gen. Rel. Grav. {\bf 54} (2022) 20.

\bibitem{PLB} S. Carneiro, P. C. de Holanda and A. Saa, Phys. Lett. {\bf B822} (2021) 136670.

\bibitem{Nelson} I. J. Araya {\it et al.}, JCAP {\bf 2302} (2023) 030.

\bibitem{bojowald} M. Bojowald, Phys.Rev. {\bf D103} (2021) 126025.

\bibitem{florencia} F. Ben\'itez, R. Gambini and J. Pullin, arXiv: 2102.09501 [gr-qc], 2021.

\bibitem{referee2} N. Bodendorfer, F. M. Mele and and J. M\"unch, Class. Quantum Grav. {\bf 36} (2019) 187001.

\bibitem{guillermo3} A. Garcia-Quismondo and G. A. Mena Marug\'an, Phys. Rev. {\bf D106} (2022) 023532.

\bibitem{referee1} S. Kazemian {\it et al.}, Class. Quantum Grav. {\bf 40} (2023) 087001.

\bibitem{bascos2} A. Alonso-Bardaji, D. Brizuela and R. Vera, Phys. Rev. {\bf D106} (2022) 024035.

\bibitem{Rovelli} C. Rovelli and E. Wilson-Ewing, Phys. Rev. {\bf D90} (2014) 023538.



\end{thebibliography}
\end{document}